\documentclass[twocolumn,twoside,slac_two]{revtex4}
\usepackage{graphicx}
\usepackage{url}
\usepackage{fancyhdr}
\usepackage{graphics}
\usepackage{epstopdf}
\usepackage{textpos}
\begin{document}

\title{\centering Measurement of the ATLAS di-muon trigger efficiency
in proton-proton collisions at 7 TeV}


\author{
\centering
\includegraphics[scale=0.06]{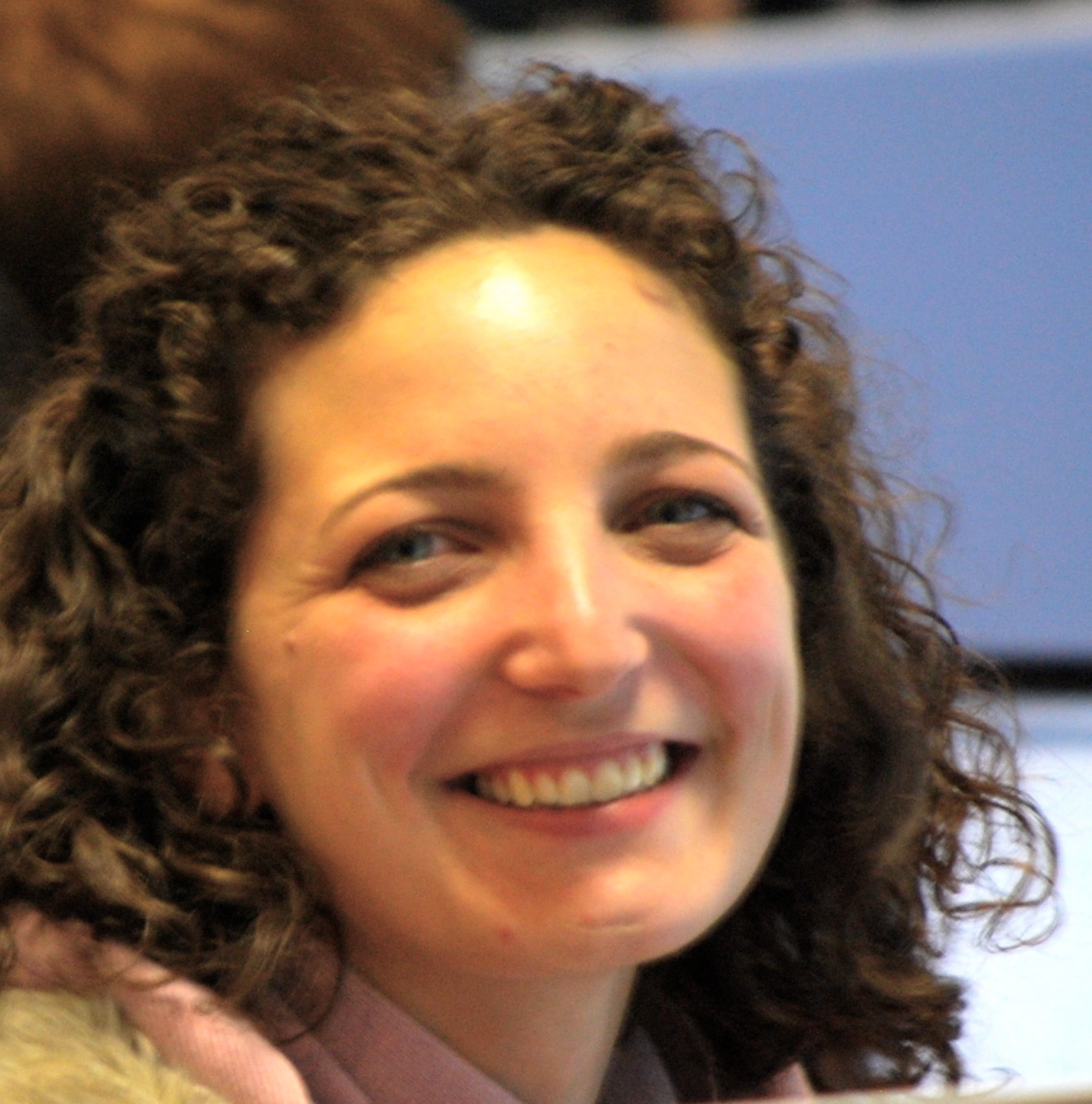} \\
\begin{center}
E.Piccaro, Queen Mary, University of London, United Kingdom
\end{center}}
\affiliation{\centering On Behalf of the ATLAS Collaboration}

\begin{abstract}
At the LHC, muons are produced in many final states and used in a variety of analysis, such
as Standard Model precision measurements and searches for new physics.
The B-physics programme in ATLAS includes the measurement of CP violating effects in B meson
decays,
the search for rare b decay signatures, as well as the study of the production cross sections.
The ATLAS detector can identify muons with high purity in a transverse momentum ($p_{T}$) 
range from a few GeV to several TeV.
In order to achieve a high trigger efficiency for low $p_{T}$ di-muon events 
and at the same time keep an acceptable trigger rate, 
dedicated trigger algorithms have been designed and implemented in the trigger menu since the 2010
data taking period.
There are two categories of B-physics triggers, one topological and one non-topological.
Both of these have been studied and their performance assessed using collision data at $\sqrt{s}$ = 7 TeV.
The performance found with data has been verified with simulated events.
\end{abstract}

\maketitle
\thispagestyle{fancy}


\section{Di-muon Trigger in ATLAS}
At the design luminosity of $10^{34}$cm$^{-2}$s$^{-1}$ the LHC's bunch crossing 
rate is 40 MHz at each of the four interaction points
(ATLAS, CMS, LHCb, ALICE). The ATLAS trigger system is designed to reduce this rate,
and record events at approximately 200Hz. In order to reduce the rate the system 
is made of three levels. The Level 1 (L1) is hardware based and uses information
from the calorimeter and the fast muon trigger detectors at reduced granularity.
L1 reduces the input rate to a maximum
of 75 kHz (30 kHz in 2010) and also identifies Regions of Interest (RoIs) that are 
then investigated by the other levels. The Level 2 (L2) and the Event Filter (EF)
are software based and use information from all the subdetectors.
Together they are known as the High Level Trigger (HLT).
  L2 reduces 
further the rate to about 3 kHz. The final reduction is done at the 
EF ~\cite{Performance}.  

To achieve this large rate reduction a $p_{T}$ threshold must be raised and interesting 
low $p_{T}$ events are lost. Therefore ATLAS has developed di-muon triggers 
with lower muon $p_{T}$ thresholds. These allow events of interest for the
B-physics community to be recorded. This choice accommodates both the rate
reduction as well as a high efficiency at the HLT.

\begin{figure}
\includegraphics[width=23mm]{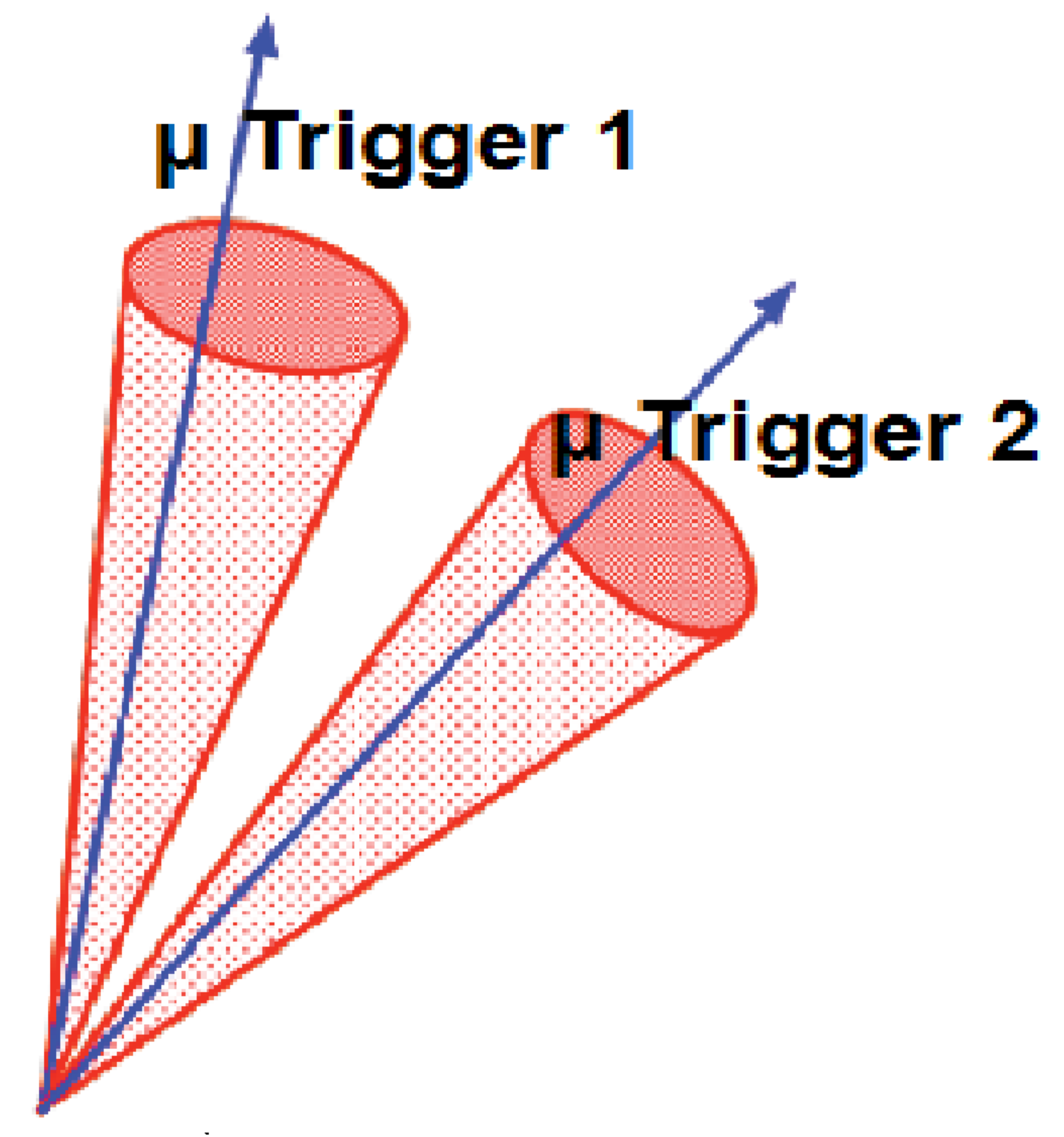}
\includegraphics[width=23mm]{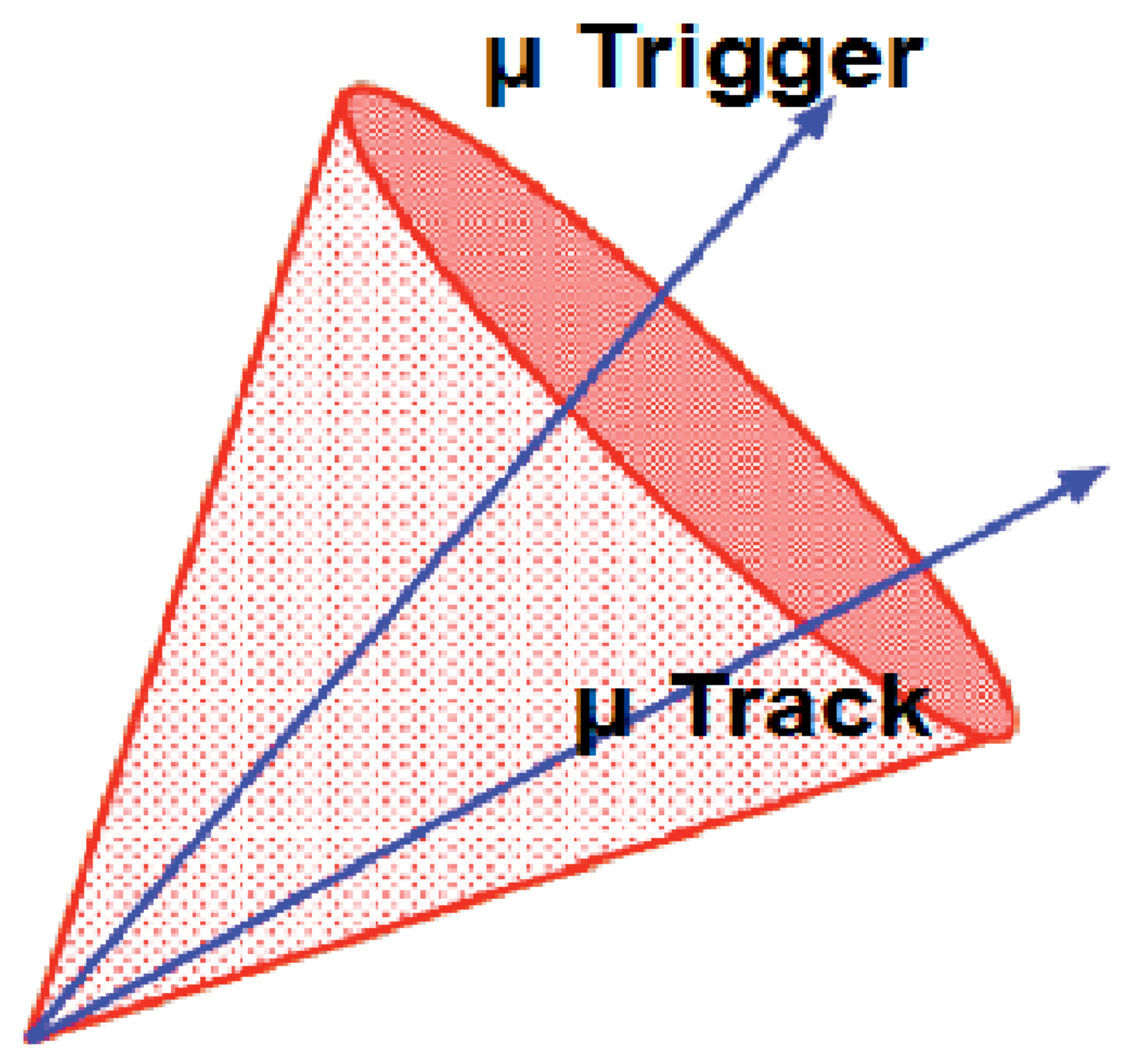}
\caption{Diagram showing how the two different B-physics trigger algorithms are
configured. The diagram on the left shows a generic topological trigger, the 
one on the right shows the TrigDiMuon trigger.}
\label{Diagrams}
\end{figure}

B-physics triggers can be divided into two categories (illustrated in Figure \ref{Diagrams}),
depending on the algorithm
used at L2 to select the di-muon pair. The TrigDiMuon triggers are seeded by a single L1 
muon RoI. The HLT algorithm then searches for a second muon within a wide $\eta$, $\phi$
region, where $\eta$ is the pseudorapidity and $\phi$ the azimuthal angle. 
This is done by extrapolating Inner Detector (ID) tracks to the muon spectrometer.
The extrapolation can also be done by looking into the whole detector, however this 
strategy is not followed in this poster. 
The second category are the topological di-muon triggers which are seeded by two 
L1 muon RoIs. Each muon is then confirmed separately at the HLT. 

Both categories of triggers apply extra requirements in invariant mass,
opposite charge and  vertex $\chi^{2}$ match in order to reduce the rate 
and yet keep interesting J/$\psi$, $\Upsilon$ and B meson data.

As an example, the di-muon triggers studied below have the {\tt \_Jpsimumu}
extension in their name. This refers to the hypothesis which in this case 
is summarised as follows:

\begin{itemize}
\item Oppositely charged muon trigger objects;
\item Invariant mass in the range from 2.5 GeV to 4.3 GeV;
\item Vertex $\chi^{2}$ match $<$ 20;
\end{itemize}

\section{Trigger efficiency measurement for di-muon triggers from 
2010 data and Monte Carlo comparison}

The accurate determination of the trigger efficiency is of great importance 
for precise cross section measurements in the low energy regime. In fact the 
$p_{T}$ spectrum of muons from   J/$\psi$ and $\Upsilon$ decays is soft and  populates the 
low trigger threshold region.

Offline, the selection of two oppositely charged combined muons with invariant 
mass in the range from 2.8 GeV to 3.34 GeV minimizes the 
background contamination and selects a pure J/$\psi$ sample. The trigger efficiencies 
determined below are measured with respect to this class of muons. 

\begin{figure}[h!]
\includegraphics[width=65mm]{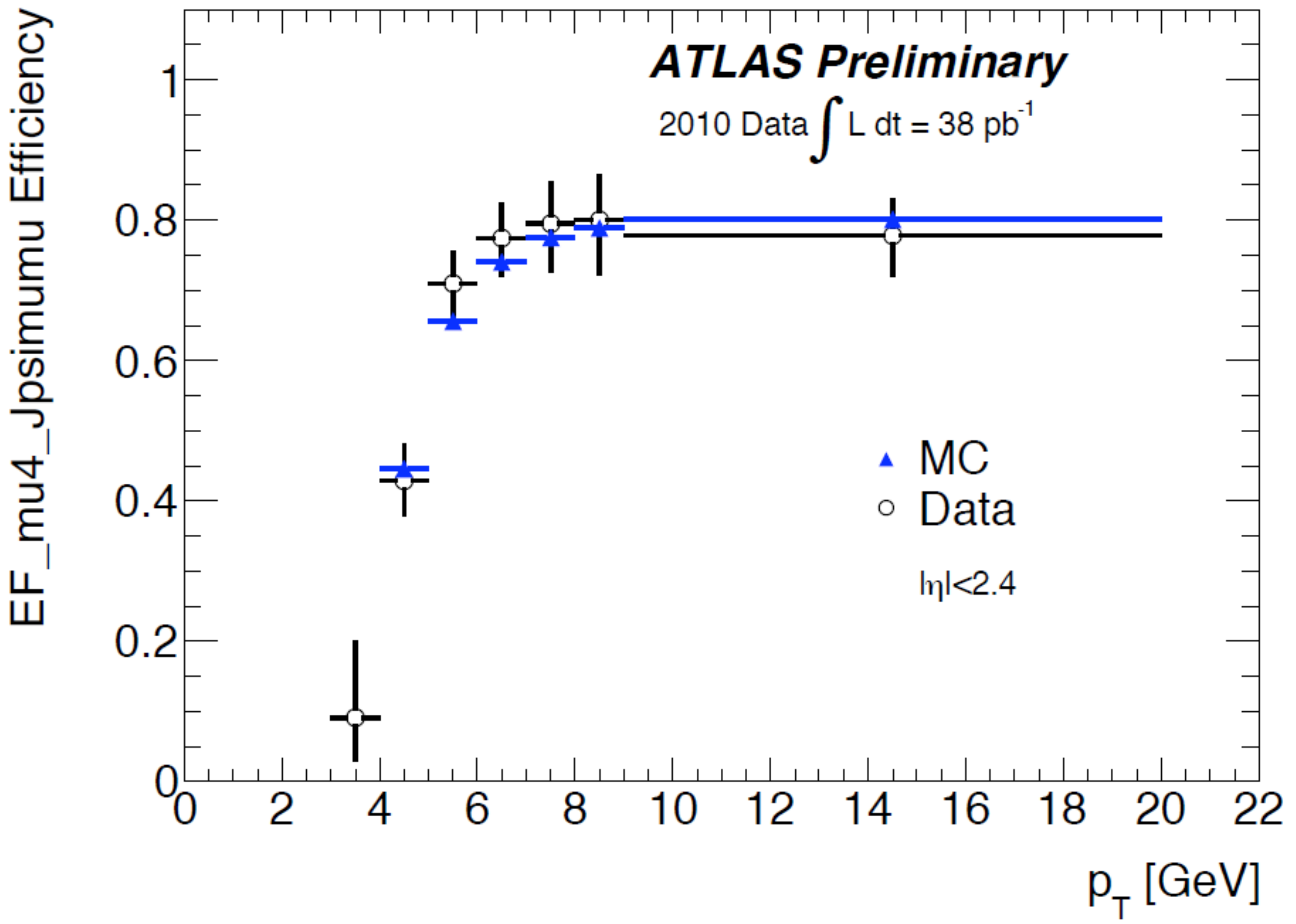}
\includegraphics[width=65mm]{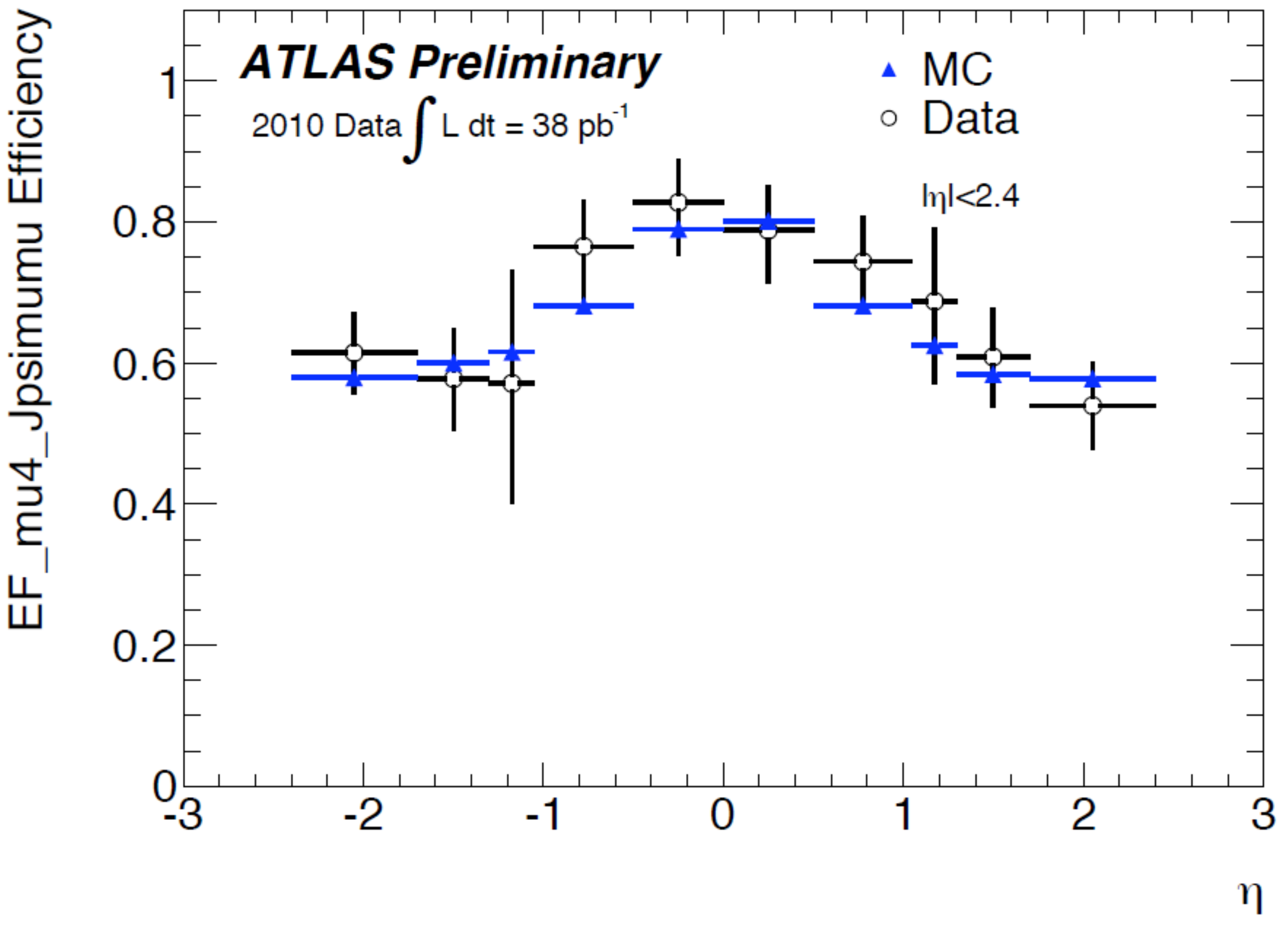}
\caption{Efficiency of {\tt EF\_mu4\_Jpsimumu} as a function of the $p_{T}$ (top) and $\eta$ (bottom)
of the reconstructed muon with higher $p_{T}$ in the di-muon pair \cite{Plots} .} 
\label{EfficiencyTDM}
\end{figure}

The method presented in the poster is based on Bayes' theorem. 
According to this, the efficiency of a di-muon trigger can be expressed as
\begin{equation}\label{eq:theorem}
\varepsilon(B)=\frac{\varepsilon(A)\varepsilon(B|A)}{\varepsilon(A|B)}
\end{equation}
where $\varepsilon(B)$ is the trigger efficiency  of the  di-muon chain and 
$\varepsilon(A)$ is the
efficiency of single muon trigger, i.e. {\tt EF\_mu4} which is an EF  single 
muon trigger passing the 4GeV transverse momentum threshold. The two extra factors $\varepsilon(B|A)$ and 
$\varepsilon(A|B)$ are the conditional efficiency terms.

The efficiency of single muon trigger has been measured in early 2010 data using the 
standard ``Tag and Probe'' method and compared to Monte Carlo simulation ~\cite{SingleMuon}.
Also, the conditional probability can be measured from data without biasing the measurement.

By combining the various studies together, the efficiencies of various
di-muon triggers have been measured using the above method in 2010 data corresponding to 
an integrated luminosity of 38 pb$^{-1}$. A good agreement with Monte Carlo simulation
has been found. 

\begin{figure}[h!]
\includegraphics[width=65mm]{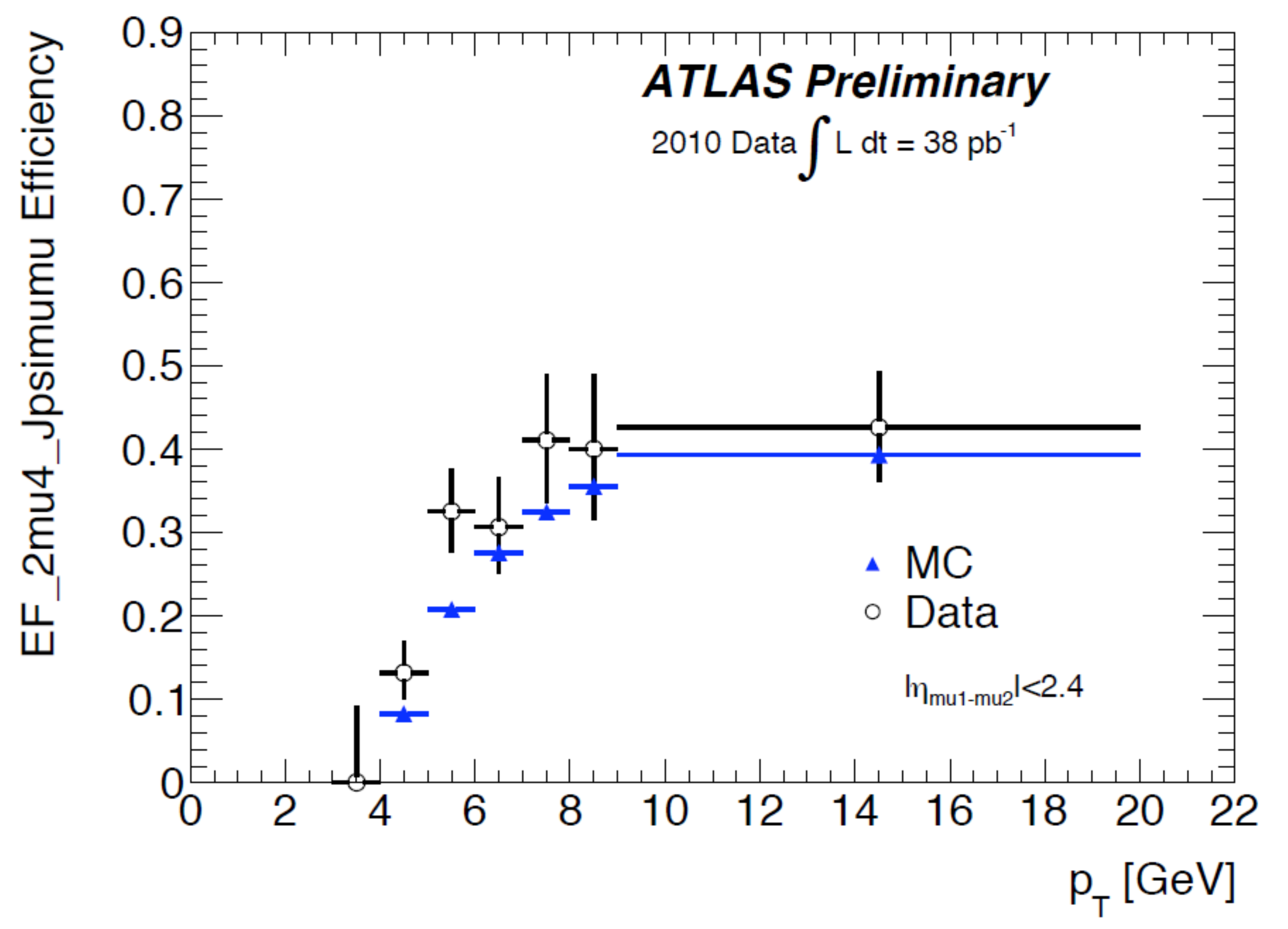}
\caption{Efficiency of {\tt EF\_2mu4\_Jpsimumu} as a function of the $p_{T}$ 
of the reconstructed muon with higher $p_{T}$ in the di-muon pair \cite{Plots}.} 
\label{EfficiencyTopo}
\end{figure}

Figure \ref{EfficiencyTDM} shows the efficiency of the TrigDiMuon chain {\tt EF\_mu4\_Jpsimumu} as a function
of the transverse momentum (top) and rapidity (bottom) of the leading reconstrunced 
muon in the di-muon pair.
An example of trigger efficiency for the topological trigger {\tt EF\_2mu4\_Jpsimumu} is shown in 
Figure \ref{EfficiencyTopo}. Here the efficiency is plotted as a function of the $p_{T}$ of the 
leading reconstrunced muon in the di-muon pair. 

By comparing the efficiency curves as a function of the transverse momentum, 
it is evident that at any given $p_{T}$, the efficiency of the topological trigger
(Figure \ref{EfficiencyTopo}) is much lower than the corresponding efficiency 
for the non-topological trigger with same hypothesis (top plot in Figure \ref{EfficiencyTDM}). 
This is due to the difference at L1 between the two trigger algorithms. Infact the topological 
trigger requires two L1 seeds, therefore the overall efficiency is proportional to the square
of the single seeded corresponding trigger.

\bigskip 
\bibliography{basename of .bib file}

\end{document}